\documentclass[twocolumn,preprintnumbers,amsmath,amssymb,floatfix]{revtex4}

\usepackage{graphicx}
\usepackage{dcolumn}
\usepackage{bm}

\begin{document}

\title{Pseudorapidity dependence of charged hadron transverse momentum spectra in d+Au collisions 
at $\sqrt{s_{_{\it NN}}} =$ 200 GeV }
\date{\today}
\author{
B.B.Back$^1$,
M.D.Baker$^2$,
M.Ballintijn$^4$,
D.S.Barton$^2$,
B.Becker$^2$,
R.R.Betts$^6$,
A.A.Bickley$^7$,
R.Bindel$^7$,
W.Busza$^4$,
A.Carroll$^2$,
M.P.Decowski$^4$,
E.Garc\'{\i}a$^6$,
T.Gburek$^3$,
N.George$^2$,
K.Gulbrandsen$^4$,
S.Gushue$^2$,
C.Halliwell$^6$,
J.Hamblen$^8$,
A.S.Harrington$^8$,
C.Henderson$^4$,
D.J.Hofman$^6$,
R.S.Hollis$^6$,
R.Ho\l y\'{n}ski$^3$,
B.Holzman$^2$,
A.Iordanova$^6$,
E.Johnson$^8$,
J.L.Kane$^4$,
N.Khan$^8$,
P.Kulinich$^4$,
C.M.Kuo$^5$,
J.W.Lee$^4$,
W.T.Lin$^5$,
S.Manly$^8$,
A.C.Mignerey$^7$,
R.Nouicer$^{2,6}$,
A.Olszewski$^3$,
R.Pak$^2$,
I.C.Park$^8$,
H.Pernegger$^4$,
C.Reed$^4$,
C.Roland$^4$,
G.Roland$^4$,
J.Sagerer$^6$,
P.Sarin$^4$,
I.Sedykh$^2$,
W.Skulski$^8$,
C.E.Smith$^6$,
P.Steinberg$^2$,
G.S.F.Stephans$^4$,
A.Sukhanov$^2$,
M.B.Tonjes$^7$,
A.Trzupek$^3$,
C.Vale$^4$,
G.J.van~Nieuwenhuizen$^4$,
R.Verdier$^4$,
G.I.Veres$^4$,
F.L.H.Wolfs$^8$,
B.Wosiek$^3$,
K.Wo\'{z}niak$^3$,
B.Wys\l ouch$^4$,
J.Zhang$^4$\\
\vspace{2mm}
(PHOBOS Collaboration) \\
\vspace{2mm}
\small
$^1$~Physics Division, Argonne National Laboratory, Argonne, IL 60439-4843,
USA\\
$^2$~Chemistry and C-A Departments, Brookhaven National Laboratory, Upton, NY
11973-5000, USA\\
$^3$~Institute of Nuclear Physics PAN, Krak\'{o}w, Poland\\
$^4$~Laboratory for Nuclear Science, Massachusetts Institute of Technology,
Cambridge, MA 02139-4307, USA\\
$^5$~Department of Physics, National Central University, Chung-Li, Taiwan\\
$^6$~Department of Physics, University of Illinois at Chicago, Chicago, IL
60607-7059, USA\\
$^7$~Department of Chemistry, University of Maryland, College Park, MD 20742,
USA\\
$^8$~Department of Physics and Astronomy, University of Rochester, Rochester,
NY 14627, USA\\
}

\begin{abstract}\noindent

We have measured the transverse momentum distributions of charged hadrons in d+Au
collisions at $\sqrt{s_{NN}}$ = 200 GeV in the range of $0.5 < p_T < 4.0$ GeV/c.  The
total range of pseudorapidity, $\eta$, is $0.2 < \eta < 1.4$, where positive
$\eta$ is in the deuteron direction.  The 
data has been divided into three regions of pseudorapidity, covering
$0.2 < \eta < 0.6$, $0.6 < \eta < 1.0$, and $1.0 < \eta < 1.4$ and 
has been compared to charged
hadron spectra from $\textrm{p}+\overline{\textrm{p}}$ collisions at the same energy.  
There is a 
significant change in the spectral shape as a function of pseudorapidity.  As $\eta$
increases we see a decrease in the nuclear modification factor $R_{\it dAu}$.

\vspace{3mm}
\noindent 
PACS numbers: 25.75.-q,25.75.Dw,25.75.Gz
\end{abstract}

\maketitle

The yields of charged hadrons from d+Au collisions at $\sqrt{s_{NN}}$ = 200 GeV
from the Relativistic Heavy Ion Collider (RHIC) have been measured by PHOBOS as a function 
of pseudorapidity, $\eta$, and transverse momentum, $p_T$.  One of the goals of the 
d+Au run was to determine whether the suppression relative
to the number of binary nucleon-nucleon collisions ($N_{\it coll}$) seen
in the high $p_T$ region in Au+Au collisions 
\cite{Phenix_quench,Phenix_AuAu_cent,Phenix_AuAu_pizero,Star_AuAu_cent,Phobos_AuAu} 
was due to initial 
state effects, such as parton saturation \cite{first_part_sat}, or due to 
final state effects of particles interacting with the dense medium produced in the 
collision \cite{jetquench}.

If supression at high $p_T$ in Au+Au data was caused by initial state effects,
a suppression would also be seen in the d+Au data from peripheral to
central collisions \cite{part_sat_npart_theory}.  The
fact that no suppression was seen \cite{Phenix_dAu,Brahms_dAu,Star_dAu,Phobos_dAu} leads 
to the conclusion that the suppression in
Au+Au data was due to interactions in the final state.

The d+Au results from midrapidity \cite{Phenix_dAu,Brahms_dAu,Star_dAu} 
show an enhancement at high $p_T$ similar to data from lower energies \cite{Cronin}, while
the results from PHOBOS \cite{Phobos_dAu}, which have an average pseudorapidity of 0.8, show 
no clear
enhancement at high $p_T$.  A more detailed study as a function of $\eta$ was
performed in order to investigate this difference in the data.  In addition, it has been
predicted that parton saturation effects, while not responsible for the
suppression seen in the mid-rapidity Au+Au data, might cause a decrease in the
particle yields at high rapidity in p+A and d+A relative to p+p 
collisions \cite{part_sat_eta}.  In this paper we average the data presented in 
\cite{Phobos_dAu} over centrality, divide it into three different bins of pseudorapidity, 
$0.2 < \eta < 0.6$, $0.6 < \eta < 1.0,$ and $1.0 < \eta < 1.4$, and
examine the yields relative to $\textrm{p}+\overline{\textrm{p}}$ collisions for each region. 

This analysis was nearly identical to that of \cite{Phobos_dAu}, with three
important changes. First, we omitted the events triggered by high $p_T$ particles
which traverse the spectrometer due to the limited $\eta$ acceptance of the 
Time-of-Flight walls.
Second, the corrections to the
spectra for geometrical acceptance, tracking efficiency, momentum
resolution and binning distortion were determined and applied separately
for each pseudorapidity bin. Finally, we averaged over centralities,
and calculated $N_{coll}$ (the average number of binary collisions) 
using a weighted average with the
number of events in each centrality bin. The mean $N_{coll}$ in these data
is $9.5 \pm 0.8$ (systematic).  The vertex range along the beam axis of the data used in this analysis is
$-10$ ${\rm{cm}} < z_{vtx} < 10$ ${\rm{cm}}$ with respect to the nominal interaction point.  Since the
$z_{vtx}$ and the accepted range of $\eta$ in the spectrometer are highly correlated,
the different $\eta$ regions correspond to different ranges in $z_{vtx}$.  The ranges
in $z_{vtx}$ for the different $\eta$ bins are shown in Table \ref{table1}.

\begin{table}[htbp]
\begin{center}
\begin{tabular}{|c|c|}
\hline $\eta$ range & range in $z_{vtx}$ [cm]\\
\hline
 $0.2 < \eta < 0.6$ & $0 < z_{vtx} < 10$ \\
 $0.6 < \eta < 1.0$ & $-10 < z_{vtx} < 5$ \\
 $1.0 < \eta < 1.4$ & $-10 < z_{vtx} < 5$ \\
\hline
\end{tabular}
\caption{
\label{table1}
Ranges in $z_{vtx}$, the position of the vertex along the beam axis, used for the 
different pseudorapidity bins.}
\end{center}  
\end{table} 

\begin{figure*}[htbp]
\includegraphics[width=15cm]{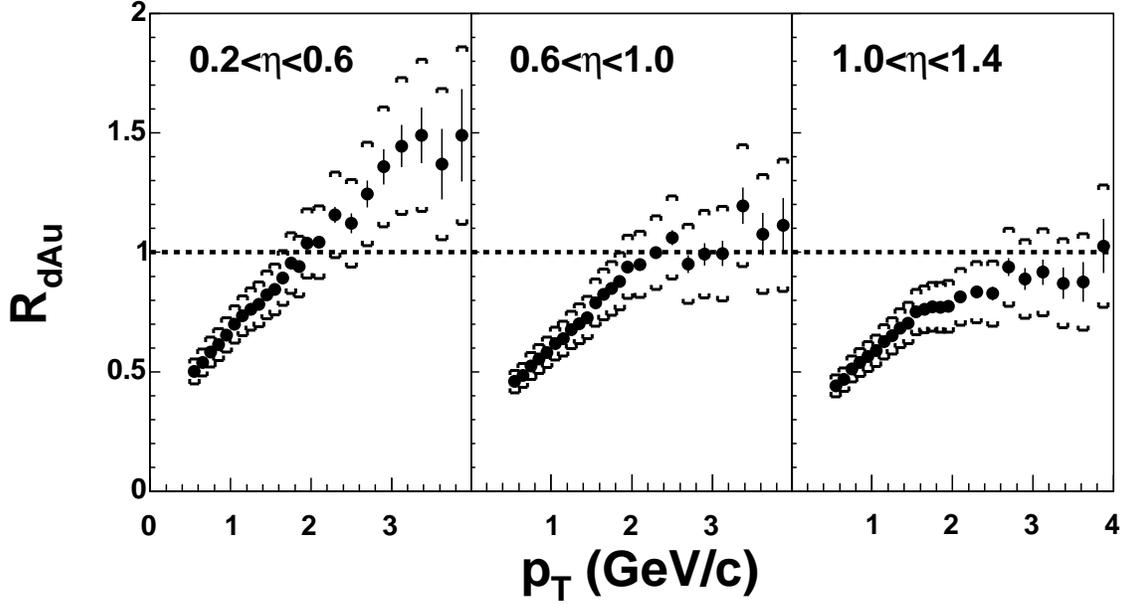}
\caption{ \label{RdAuEtaBins} 
Nuclear modification factor $R_{dAu}$ for three different pseudorapidity ranges.  The
brackets show the systematic uncertainty (90\% C.L.).}
\end{figure*}

In Figure \ref{RdAuEtaBins} we show the nuclear modification factor $R_{\it dAu}$ for
charged hadrons, defined as 

\begin{equation}
R_{\it dAu} = \frac{\sigma_{p\overline{p}}^{inel}}{\langle N_{\it coll} \rangle} 
              \frac{d^2 N_{\it dAu}/dp_T d\eta} {d^2 \sigma(\mbox{UA1})_{p\overline{p}}/dp_T d\eta},
\end{equation}
for the three different $\eta$ bins.  The $\textrm{p}+\overline{\textrm{p}}$ reference 
data used is
from UA1 at 200 GeV \cite{UA1}.  Since the UA1 acceptance ($|\eta| < 2.5$) is
quite different from the PHOBOS spectrometer acceptance, we applied a correction to 
the UA1 spectra \cite{Phobos_dAu}, which was determined using
the PYTHIA event generator \cite{pythia}.  A separate correction to the UA1 data was generated
for each of the different $\eta$ bins.  The decrease seen in $R_{\it dAu}$ with increasing 
pseudorapidity is qualitatively consistent with the predictions of the parton saturation 
model \cite{part_sat_eta}, as well as with the data from the BRAHMS experiment
\cite{Brahms_eta}, which extend to $\eta$ = 3.2.

\begin{figure}[htbp]
\includegraphics[width=8cm]{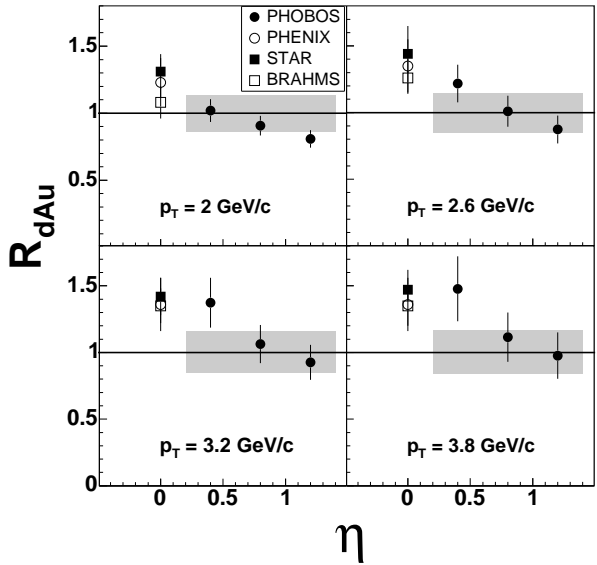}
\caption{ \label{RdAuVsEta} 
Nuclear modification factor $R_{\it dAu}$ for four different values
of $p_T$ for the three different $\eta$ bins.  The error bands show the
common systematic scale errors (90\% C.L.)
for the PHOBOS data.  The error bars show the systematic
point to point uncertainty (90\% C.L.) of the fitted data points.
PHENIX, STAR, and BRAHMS points are from Refs.
\cite{Phenix_dAu,Brahms_dAu,Star_dAu}.}
\end{figure}

To further understand the evolution of $R_{\it dAu}$ with pseudorapidity we have plotted the
values of $R_{\it dAu}$ for different $p_T$ values as a function of $\eta$, shown in Figure
\ref{RdAuVsEta}.
A function was fit to the spectra which consisted of the sum of an exponential and a
power law.  This fit was then divided by a fit to the corrected UA1 data.  The ratio
of the values of these two fit functions are plotted at four different values of $p_T$ as
a function of $\eta$.  We see a smooth decrease in $R_{\it dAu}$ from 
mid to high pseudorapidity.  The gray error bands represent the correlated
scale errors for the different $\eta$ bins at the given $p_T$.  The main 
contributions to these errors are the error on $N_{coll}$ and the global uncertainty in
the acceptance and efficiency correction.  The error bars show the uncorrelated
point to point errors between the $\eta$ bins.  These are dominated by
the uncertainty in the $\eta$ dependence of the acceptance and efficiency corrections, 
the dead channel correction, and the momentum resolution corrections.

In summary, we have measured the pseudorapidity dependence of the nuclear
modification factor derived from the yield of charged hadrons produced in d+Au
collisions at $\sqrt{s_{_{\it NN}}} =$ 200 GeV. 
A smooth decrease of $R_{\it dAu}$ with increasing
pseudorapidity has been observed. 
The details of the
evolution of $R_{\it dAu}$ as a function of transverse momentum and pseudorapidity
can be used to evaluate and constrain models including saturation effects.

This work was partially supported by US DoE grants DE-AC02-98CH10886,
DE-FG02-93ER40802, DE-FC02-94ER40818, DE-FG02-94ER40865,
DE-FG02-99ER41099, W-31-109-ENG-38, US NSF grants 9603486, 9722606,
0072204, Polish KBN grant 2-P03B-10323, and NSC of Taiwan contract NSC
89-2112-M-008-024.

\end{document}